# Exploring a Virtual Pet to Provide Context Notifications in a Tourism Recommender System: a Pilot Study

**Patrícia Alves**
(GECAD, ISEP, Polytechnic of Porto, Porto, Portugal / LASI, University of Minho, Braga, Portugal
https://orcid.org/0000-0003-3997-311X, pat@isep.ipp.pt)

**Joana Neto**
(ISEP, Polytechnic of Porto, Porto, Portugal
https://orcid.org/0009-0009-4406-2952, jmasn@isep.ipp.pt)

**Ana Barreiro**
(ISEP, Polytechnic of Porto, Porto, Portugal
https://orcid.org/0009-0000-0224-5632, 1190353@isep.ipp.pt)

**Jorge Lima**
(ISEP, Polytechnic of Porto, Porto, Portugal
https://orcid.org/0009-0001-6600-6857, 1210819@isep.ipp.pt)

**Fausto Alves**
(ISEP, Polytechnic of Porto, Porto, Portugal
rfmas@isep.ipp.pt)

**Henish Balu**
(Porto Digital, Porto, Portugal
henish.balu@portodigital.pt>)

**Luís Conceição**
(GECAD, ISEP, Polytechnic of Porto, Porto, Portugal / LASI, University of Minho, Braga, Portugal
https://orcid.org/0000-0003-3454-4615, msc@isep.ipp.pt)

**Goreti Marreiros**
(GECAD, ISEP, Polytechnic of Porto, Porto, Portugal / LASI, University of Minho, Braga, Portugal
https://orcid.org/0000-0003-4417-8401, mgt@isep.ipp.pt)

**Abstract:** While context-aware personalization has been widely explored in modern tourism Recommender Systems (RS), the delivery of real-time notifications remains a significant design challenge due to issues of intrusiveness and user fatigue. This paper presents a proof-of-concept for a tourism recommendation framework that utilizes a virtual pet as a social mediator for delivering context-aware alerts. The system integrates real-time environmental data - including air quality, noise levels, and weather forecasts - and proximity-based notifications with a Multi-Agent Microservice that generates personalized recommendations based on the user's personality traits and preferences.
A within-subjects pilot study ($n = 11$) was conducted to evaluate the feasibility and user acceptance of this pet-mediated approach. Participants interacted with two versions of the system - a baseline without contextual alerts and a version featuring pet-mediated notifications - over a four-week period (two weeks per version) in real-world scenarios. Quantitative and qualitative data were collected to assess engagement, perceived naturalness, notification utility, and acceptance. Preliminary results suggest that the virtual pet effectively can "soften" the perceived intrusiveness of system alerts, making safety-critical information feel more welcome and natural. Furthermore, the character-mediated justifications significantly improved the clarity of the notifications, effectively supporting users in their real-time travel decisions. These findings provide a foundation for using virtual pet companions to enhance the transparency and acceptance of context-aware communication in tourism RS.

**Keywords:** Virtual Pet, Recommender System, Context notification, Gamification, Context-Awareness, Leisure Tourism
**Categories:** H.3.3, H.3.7, H.4.3, H.5.1, J.4

## 1 Introduction

A few decades ago, tourism information was primarily retrieved through general search engines, often leading to information overload and choices poorly aligned with a traveler's actual desires. The advent of Recommender Systems (RS) and mobile technology has significantly improved this process, allowing for real-time, context-aware suggestions. However, a persistent challenge remains: the intrusiveness of notifications. When alerts are delivered without considering the user's temporal, spatial, or environmental context - such as suggesting outdoor activities during poor weather or air quality - they risk being perceived as bothersome, leading to "notification fatigue" and system abandonment. For example, when a tourist is prompted by the app to visit an outdoor point of interest (POI) but the weather conditions are unfavorable, or the air quality is prejudicial at the time defined for the visit.

To address these challenges, we developed a proof-of-concept for a context-aware notification framework within a previously developed tourism recommender system [Alves et al. 2024]. This system evaluates real-time adverse conditions, such as bad weather, prejudicial noise, or poor air quality, to alert the user. The system integrates a virtual assistant represented as a pet, designed to deliver these notifications in a friendly and less intrusive manner. By leveraging the social and emotional cues associated with a virtual companion, the design seeks to establish empathy and a welcoming interaction environment, shifting the perception of notifications from "system interruptions" to "helpful companion advice." It also detects if the user is walking outside in order to provide recommendations of nearby POI related to the user's preferences. The system also includes a built-in game featuring the virtual pet as the main character, where users are able to interact with the pet by providing care and participating in mini-games, further detailed in another work [Alves et al. 2025].

This paper presents the design of this pet-mediated notification system and reports on a within-subjects pilot study ($n$ = 11). The study was designed as a preliminary exploration to assess the feasibility and user acceptance of pet-mediated alerts compared to standard system notifications. To ensure a realistic interaction period, participants explored each version of the system for two weeks. Although the underlying architecture supports group contexts, this pilot study focuses on individual user perception and engagement to establish a baseline for pet-mediated communication.

Preliminary results suggest that the pet-mediated approach effectively can enhance the acceptance and perceived naturalness of context-aware alerts. Participants reported higher perceived enjoyment and a clearer understanding of notification justifications, showing a greater intention to act upon safety-critical information - such as weather and air quality warnings - when presented through the virtual companion. These findings provide initial support for the use of virtual pets as a strategy to mitigate notification intrusiveness and foster sustained engagement in mobile tourism applications.

The rest of the paper is divided into four main sections: Section 2 presents some background and related work on virtual assistants, virtual pets and context notifications in RS. Section 3 explains the methodology used to develop the context notifications using the virtual pet and to conduct the pilot study. Section 4 presents the results of the pilot study and their discussion, and, finally, Section 5 gives some final conclusions and future directions.

## 2 Background and related work

This section presents some background on virtual assistants in general and in the pet form, as well as their usage in RS, focusing on context notifications.

### 2.1 Virtual assistants

Virtual assistants have become an important cornerstone for human-computer interaction. It has been demonstrated that integrating virtual cognitive assistants can significantly enhance user empathy and motivate system adoption [Archita et al. 2025, Güell et al. 2020]. An example of the use of virtual assistants can be found in the work of [Zhao et al. 2020], who developed a personalized fitness RS that employs a digital assistant to assign a series of tasks to users. These tasks are designed to infer the users' preferences, thereby shaping the recommendations into what they prefer.

[Güell et al. 2020] conducted a study on a critique-based RS that uses a cognitive virtual assistant in order to personalize recommendations in a pleasant way. In the high-value product domain, items are considerably more expensive, and consumers are often buying them for the first time. The virtual assistant guides the user to their desired choice, improving their experience.

[Benedetto et al. 2019] proposed a Personalized Virtual Teaching Assistant (PVTA) aimed at supporting assisted learning by offering services such as content personalization, learning material recommendations, and student engagement. They also present an early version of the PVTA, implemented as a chatbot built with IBM's Watson Assistant, which can respond to students' questions regarding the content, structure, and organization of the RecSys course, an introductory class on RS.

More recently, the focus has shifted toward making these assistants less disruptive. While mainstream assistants like Siri or Alexa are widely adopted, they are often criticized for their potential intrusiveness, particularly when delivering spoken notifications during focused activities, such as when listening to music. [Wang et al. 2024] addressed this by designing a virtual assistant that synchronizes notifications with the rhythm of the user's music (like directions when driving a car, exercising, or shopping), aiming for a more seamless integration.

Another example of virtual assistants is the medical cognitive assistant proposed by [Weerasinghe et al. 2024]. The system is called CognitiveEMS and is presented as an end-to-end wearable cognitive assistant, via Augmented Reality smart glasses, that operates as a virtual partner of the user. The system enables real-time collection and analysis of multimodal data in emergency scenarios and supports emergency medical services' responders.

Anthropomorphic design has shown to boost user engagement by up to 35% compared to traditional notifications [Vasalou et al. 2016, Simas and Ulbricht 2024], maintaining long-term user adherence to intervention protocols in health through a sense of "therapeutic alliance" or companionship [Bickmore and Picard 2005, Betty et al. 2022, Burger et al. 2022, Six et al. 2025].

These examples illustrate the diversity of virtual assistants but also highlight a common challenge: the balance between being helpful and being intrusive. While these systems are functional, they often lack an affective or playful dimension that could further soften the delivery of context-aware alerts. Our work builds on this by exploring how a virtual pet companion - rather than a generic anthropomorphic agent - can mediate system-to-user communication, potentially making notifications feel more natural and welcome.

### 2.2 Virtual pets

Games have exerted a substantial influence on society, particularly in supporting the cognitive, emotional, and social development of children and adolescents [Franceschini et al. 2022, Alborzi and Torabi 2024, Pitic and Pitic 2022]. Playful environments have been shown to contribute positively to emotional regulation, anxiety management, coping with loss, personal growth, and overall well-being [Gottman 1986, McDaniel and Vick 2010]. Within the digital domain, video games have sometimes been criticized for potential negative impacts on mental health and for allegedly encouraging intellectual passivity [Granic et al. 2014]. Nevertheless, growing empirical evidence suggests that well-designed digital games can foster a wide range of cognitive, motivational, and problem-solving skills, comparable to those promoted by traditional forms of play such as board games, card games, educational games, and sports.

These positive effects have motivated the adoption of game-inspired design principles in non-entertainment contexts through gamification [Deterding et al. 2011]. By integrating mechanics such as rewards, progression systems, and interactive narratives, gamification has demonstrated its potential to increase user engagement, sustain motivation, and encourage long-term interaction with digital platforms [Hamari et al. 2014]. Virtual pets represent a particularly effective gamification strategy, as they leverage emotional attachment, empathy, nurturing behaviors, and continuous interaction to foster user commitment and engagement [Thirumaran et al. 2021]. The emotional bond formed between users and virtual pets can encourage repeated system usage and promote active participation in system-driven activities. An interesting study on virtual pets is the review performed by Dodd, Fowler and Lottridge [Dodd et al. 2025], where they present their usage through different six perspectives aiming to understand their cultural significance: health and habits, life and death, gender, ethics, toys and play, and capitalism and consumption.

Virtual pets are essentially virtual characters in the form of animals, real or from fantasy, like dragons or unicorns, used to promote a sense of care and empathy towards the user by interacting with them. One of the first virtual pet was Tamagotchi, a virtual pet game released in 1996. Before smartphones were invented, the only way for a person to take care of a handheld virtual pet in the 1990s was through the egg-like device that contained it [Lawton 2017, Bylieva et al. 2019], which could be taken everywhere and take care of at anytime. It's nurturing gameplay mechanics, requiring continuous care (feeding, cleaning, and attention), strengthened users' emotional attachment and sense of responsibility. It also marked the transition of virtual pets from computer-based systems to widely accessible consumer devices, accelerating global adoption of the concept.

Another example of a virtual pet is seen in the game Nintendogs released in 2005 by Nintendo. The objective of the game was to take care of one or various virtual dogs by interacting with them through the mobile device Nintendo DS [Sánchez-Moreno 2025, Bylieva et al. 2019]. The game leverages hardware features such as touchscreens and microphones to allow players to pet, train, and communicate with the virtual animals, creating a more immersive and natural interaction model. It reflects a trend toward increasing simulation fidelity and more intuitive user interfaces in virtual pet development. Given the nature and portability of it, children began to play it intensively, creating a significant attachment towards their companions.

My Talking Tom is an example of the mobile and entertainment-oriented generation of virtual pets [Bylieva et al. 2019]. The game focuses on accessibility, frequent interaction, and playful engagement, including speech imitation and mini-games. This platform illustrates how modern virtual pets prioritize entertainment, personalization, and continuous interaction through mobile technologies.

Perhaps one of the most notable applications of virtual pets is Pou, a game released in 2012 that focuses on caring for an alien creature, conceptually similar to the previously mentioned Tamagotchi. In 2014, it was recognized as the most downloaded game [Glenday and Heinlin 2016, Wolf 2021]. At the moment, according to the Google Play Store (Android's digital marketplace), the game has surpassed 1 billion downloads, demonstrating its significant influence within its domain [Google Play 2026].

In addition to entertainment, virtual pets have been increasingly adopted into specialized domains such as health [Ivezić et al. 2024, Na et al. 2022], education [Al Hakim et al. 2024, Pujiastuti et al. 2025], and social networking [Lin and Zaborowski 2025]. Their effectiveness in these areas stems from their ability to serve as social proxies - digital entities that users perceive with a degree of social presence and empathy.

While traditional virtual pets (e.g., Pou or Tamagotchi) focus on nurturing as the primary gameplay loop, a more recent application involves using the pet as a communication mediator. In this role, the pet moves beyond a mere entertainment element to become a "companion agent" that delivers system information. This paradigm is particularly relevant for notification design. By delivering alerts through a virtual pet, systems can leverage the user's emotional bond and empathy to mitigate the perceived intrusiveness of the message. Instead of a cold, automated system alert, the pet provides "friendly advice," potentially increasing the user's willingness to accept and act upon the information. This is what we sought to explore in this proof-of-concept. Subsection 2.3 discusses how this companion-based communication is being integrated into Recommender Systems.

### 2.3 Virtual pets in recommender systems

While RS are essential for filtering information overload [Roy and Dutta 2022], recommendation content is often delivered through generic, weakly personalized notifications, leading to low engagement and high dismissal behavior [Sahami et al. 2014, Mehrotra et al. 2015, Wang et al. 2024]. Virtual pets offer a unique solution to this "notification fatigue" by serving as emotionally engaging mediators. Beyond encouraging interaction, pets can facilitate the implicit acquisition of preferences and, crucially, tailor the delivery of context-aware alerts to the user's immediate situation. This suggests that emotionally compelling pet-mediated communication can significantly improve the acceptance of notifications, especially in systems requiring continuous user input.

Beyond the general applications of virtual assistants in RS mentioned in Subsection 2.1, there is limited research exploring how the likability of pets as assistants can be used to create a better experience for RS users, especially in tourism related RS. Among the few studies found, Lin et al. [Lin and Zaborowski 2025] conducted research about a short-video RS app called Douyin, which is similar to TikTok. This app implements various forms of gamification techniques, such as friendship badges, user levels, and a virtual pet. The latter is implemented in order for users to interact tightly every day by sending texts and videos to each other, and is shown to successfully do so.

Ivezić et al. [Ivezić et al. 2024] present a health RS called EghiFit that implements a virtual pet to improve the efficiency of the app itself. The RS incentivizes users to achieve the goals agreed by their health coach by giving points that can later be spent on their pet.

Thirumaran et al. [Thirumaran et al. 2021] conducted a study with 539 participants to analyze how interactions with a virtual pet influence travel intentions and user engagement. They demonstrate that virtual pets can be effective gamification and persuasion tools capable of influencing user behavior in tourism contexts. They can be used to create emotional bonds, deliver rewarding interactions, support long-term engagement, and act as companions that guide user decisions. The authors insights strongly support the integration of virtual pets into RS, particularly in domains where experience, exploration, and user involvement are critical, such as the tourism domain.

Zhou et al. [Zhou et al. 2024] further emphasize the importance of the virtual pet's visual design, demonstrating that "cuteness" significantly promotes prosocial behavior and online cooperation. This effect is mediated by parasocial interaction, where users treat the cute virtual entity as a social partner. In the context of our work, this suggests that a visually appealing and "cute" virtual pet can not only reduce the perceived intrusiveness of notifications but also increase the user's willingness to follow system advice, effectively humanizing the interaction loop.

Our proof-of-concept extends these findings by using the virtual pet not just as a rewarding element, but as a context-aware travel companion. The pet mediates the delivery of personalized contextual recommendations and environmental alerts (e.g., weather or air quality) and provides a dedicated environment to interact with - the user can take care of pet and even play mini-games with it (Figure 1c and Figure 1d, respectively, further detailed in [Alves et al. 2025]). By leveraging the user's empathy toward the pet, the system aims to transform intrusive system alerts into "companion advice," making notifications feel more natural and welcome, justified, and useful in real-time tourism contexts.

### 2.4 Context notifications

Context-aware computing aims to give users the right information at the right time taking into account where they are, what they are doing, and what is happening around them [Dey 2001]. For RS, this means that mobile notifications need to adapt to what is actually going on in the user's life [Adomavicius and Tuzhilin 2011]. Research efforts focused at improving notification acceptance have explored multiple contextual dimensions. Temporal optimization approaches seek to identify opportune delivery moments based on user availability and interruption patterns [Pielot et al. 2014]. Attentional state modeling aims to minimize disruption during cognitively demanding activities [Fischer et al. 2010]. Location-based filtering ensures spatial relevance by adapting notification content to users' geographic context [Junglas and Watson 2008]. Social context awareness considers interpersonal situations to avoid inappropriate interruptions [Teevan and Iqbal 2016].

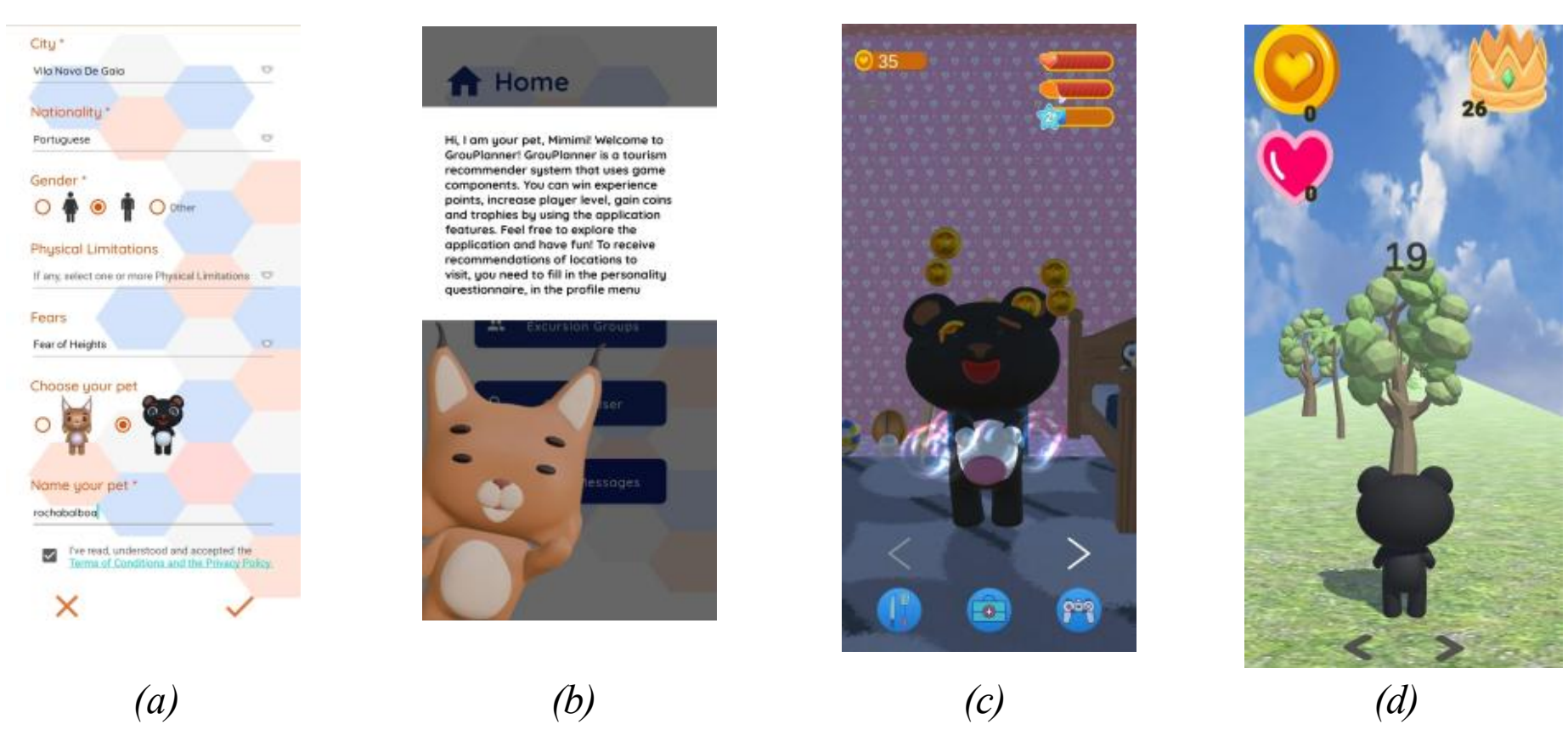

*Figure 1: Screenshots of the GRS involving the virtual pet: (a) Pet choice in the registration form; (b) Welcome screen; (c) Virtual Pet Game main scene showing the pet bathing and coins being awarded; (d) DinoRush, one of the Virtual Pet Game mini-games.*

These approaches demonstrate that multidimensional contextual modeling significantly improves notification utility and user receptiveness.

While mobile notifications remain useful, their effectiveness depends heavily on relevance, timing, personalization, and frequency [Paradeda et al. 2025, Kim et al. 2016, Jannach and Zanker 2020]. Recent studies still report notification fatigue and reduced receptiveness to intrusive or poorly timed messages, suggesting that low acceptance is still a current design challenge rather than a fully solved problem [Paradeda et al. 2025, Wang et al. 2024]. This issue is especially relevant in tourism applications, where notifications may support time-sensitive services such as weather warnings, location-based alerts, or travel updates. Because these messages compete with users' attention, poorly calibrated notification strategies may reduce engagement and lead users to ignore or disable important system communications. Accordingly, recent work supports the use of context-aware and adaptive notification strategies to improve perceived relevance and reduce intrusiveness [Wang et al. 2024].

However, while context-aware systems excel at identifying the "right moment" and "right content", there is still a gap in exploring the right channel for delivery. Most systems rely on standard OS-level alerts, which lack personal or emotional resonance. Our work addresses this by exploring the synergy between environmental context-awareness (weather, noise, air quality) and pet-mediated delivery. By using a virtual pet to communicate contextual shifts, we hypothesize that the notification's utility - already optimized by context models - can be further enhanced by the social presence and perceived "cuteness" of the companion, leading to higher trust and acceptance rates during real-time tourism exploration.

#### 2.4.1 IoT sensor integration and smart city infrastructure

Smart-city initiatives increasingly deploy distributed sensor networks and IoT infrastructures to collect real-time environmental data, including air quality, noise, weather, and other urban context signals [Torabi et al. 2023, Gonzalez et al. 2024, Anjum et al. 2025, Narayana et al. 2024]. Middleware platforms such as FIWARE support the integration and management of heterogeneous data streams through standardized APIs and context-broker architectures [FIWARE 2020, Nguyen and Nguyen 2026]. In tourism, context-aware and interactive RS have long recognized the value of contextual factors such as location, weather, time, and social setting for supporting choice-making [Jannach and Zanker 2020]. However, the use of real-time environmental data in tourism RS remains comparatively limited, especially when the goal is to support personalized or group-oriented choice-making.

Prior tourism recommender studies have shown that sensor-derived information can be used to filter or enrich suggestions, for example by adapting recommendations to weather conditions or environmental quality [Jannach and Zanker 2020, Benfares et al. 2016]. Schieder and Glatz [Schieder and Glatz 2012] found a way to filter out activities when the weather is bad. Bao et al. [Bao et al. 2015] made a system that tells tourists about the air quality in cities, although it does not take into account the individual needs of each person. More recent work has advanced these approaches by integrating real-time IoT data streams and personalized filtering mechanisms [Gonzalez et al. 2024, Nguyen and Nguyen 2026]. These approaches demonstrate the potential of contextual sensing for travel decision support, but they often rely on relatively coarse rules or limited personalization. As a result, they do not fully capture individual preferences or the potentially conflicting needs of multiple travelers. This opens an opportunity for RS that combine real-time environmental sensing with richer personalization and group-aware decision support.

In the proposed proof-of-concept, incorporating environmental information into the RS supports more adaptive and context-aware interactions. Contextual notifications, triggered by adverse weather or degraded air quality, prompt users to reconsider planned activities or explore alternative POI. These contextual cues are delivered through a gamified interface mediated by a virtual pet companion, which communicates relevant environmental information while maintaining user engagement and trust.

The present work represents a novel synthesis of these research directions in tourism RS, by combining real-time IoT environmental monitoring with a Multi-Agent Microservice (MAMS) architecture, where an individual agent is modeled for each user profile to perform personalized recommendations and notifications. This allows the system to determine if a specific environmental condition (e.g., moderate noise) is prejudicial based on that specific user's preferences or personality-based filtering (more details in [Alves et al. 2024, Alves et al. 2023]). Finally, these technical assessments are translated into emotionally compelling notifications through the virtual pet mediator. To the best of our knowledge, this is the first study to methodically combine real-time IoT sensing, individualized agent-based assessment, and anthropomorphic interface design to improve the acceptance of context-aware tourism notifications.

### 2.5 Research Gap

Virtual assistants have increasingly been incorporated into interactive systems to provide guidance and support more natural forms of interaction [Archita et al. 2025, Pardo et al. 2025, Ionitua et al. 2024]. In tourism, virtual agents have been shown to enhance user experience by delivering information in a conversational and socially engaging manner [Ferreira et al. 2023, Alamoudi et al. 2024, Melo et al. 2025]. Nevertheless, their integration within recommendation interfaces, particularly as mediators for context-aware notifications, remains limited. While context-aware systems excel at identifying relevant delivery moments based on environmental data (e.g., weather or air quality), the interaction paradigm used to deliver these alerts often relies on standard system notifications, which are frequently perceived as intrusive or impersonal.

To the best of our knowledge, the use of a virtual pet as a personalized intermediary for delivering contextual notifications in a tourism RS has not yet been explored. This work addresses this gap by proposing a context-aware notification mechanism mediated by a virtual pet assistant. The system leverages real-time environmental data and individual agent-based assessments to generate timely notifications, presenting them through the pet in a socially expressive manner. By embedding the virtual pet within the interaction flow, the system aims to transform system alerts into companion advice, enhancing situational awareness while fostering sustained engagement.

To obtain initial empirical insights, a within-subjects pilot study was conducted to assess the feasibility of this pet-mediated interface. Specifically, the study examines its impact on user engagement, the perceived naturalness and usefulness of contextual notifications, and the overall interaction experience. This exploratory evaluation aims to provide design guidelines for future large-scale implementations of social-mediated communication in tourism-oriented recommendation frameworks.

## 3 Methodology

The developed tourism recommendation architecture comprises a frontend Android client application and a backend consisting of five microservices implemented in .NET. These include the already mentioned Multi-Agent Microservice, which contains agents modeled with each tourist's profile; and a Recommendation Engine Microservice (REMS), which identifies the most suitable POI, based on the tourists' attraction preferences, travel-related preferences & concerns, disabilities, and fears/phobias, while also considering the constraints provided by the MAMS. The five microservices are exposed through a RESTful API over HTTPS, enabling interoperability with any compatible client. To address the cold-start problem commonly observed in RS, the system generates initial recommendations based on the tourists' personality traits, which are posteriorly enhanced by rules extracted by the MAMS. This approach is further enhanced by a novel personality-based clustering method designed to handle user heterogeneity and reconcile conflicting preferences within excursion groups. Full details on the underlying GRS architecture can be consulted in [Alves et al. 2024].

This work extends the framework by introducing an engagement-oriented interaction layer: a virtual pet companion acting as a gamified mediator for context-aware notifications. This layer delivers alerts regarding adverse weather and environmental conditions (noise levels and air quality), providing justifications for replacing outdoor activities and offering proximity-based POI suggestions.

The integration of the virtual pet aims to transform the delivery of contextual data into a more immersive and empathetic experience. By providing concise and intuitive recommendation justifications through character-mediated communication, the approach seeks to enhance the perceived usefulness of alerts while minimizing cognitive load. This proof-of-concept combines personalized recommendation strategies with entertainment-driven design to foster individual user engagement and assess the feasibility of social-mediated communication in tourism planning scenarios.

### 3.1 Conceptualization

The virtual pet-mediated notifications were developed to address a persistent challenge in mobile tourism: low engagement and high dismissal rates for contextual alerts [Sahami et al. 2014, Mehrotra et al. 2015, Wang et al. 2024, Dai et al. 2024, Yang et al. 2024]. As a proof-of-concept, this work explores whether a personalized virtual mascot can serve as a social mediator for contextual tourism recommendations, with three primary design objectives: (1) making technical alerts about weather, air quality, and noise more approachable through character-mediated delivery; (2) differentiating safety or comfort relevant alerts from generic push notifications to improve perceived acceptability; and (3) providing concise, character-driven explanations (e.g., suggesting an indoor POI due to rain) to support intuitive decision-making.

The notification system implements a three-layer architecture that integrates context detection, intelligent filtering, and personalized delivery (see Figure 2):

- **Context Detection Layer:** Monitors real-time environmental data - including weather forecasts, air quality, and noise levels - alongside user movement and activity via the Android Activity Recognition API.
- **Intelligent Filtering Layer:** Interfaces with the MAMS and REMS to assess the severity of environmental shifts based on the user's profile, location (if walking), or planned itinerary. This layer determines the necessity and timing of each alert.
- **Notifications Delivery Layer:** Employs a dual-channel strategy. Remote push notifications (via Firebase Cloud Messaging) are used for planned excursion updates, while local push notifications are triggered by proximity during exploration. In this proof-of-concept, the virtual pet serves as the primary interface for in-app interactive popups, providing the character-mediated communication and justifications that differentiate these alerts from standard OS-level notifications.

### 3.2 Design of the virtual pets

The virtual pets were developed using Blender, following a standard 3D character pipeline including base mesh construction, refinement of anatomical features, UV mapping, and rigging. UV mapping allowed the three-dimensional mesh to be flattened into a two-dimensional representation for texture application, enabling controlled manipulation of organic surfaces. To allow for user personalization, two distinct mascots - a panda and a lynx - were created (see Figure 1a).

A key design requirement for this proof-of-concept was the optimization of "cuteness" to foster user empathy and acceptance. Cuteness, often defined as features that evoke feelings of warmth and protectiveness [Septianto and Paramita 2021], has been shown to significantly promote prosocial behavior and online cooperation [Zhou et al. 2024]. Consequently, the base mesh utilized primitive geometric shapes refined into a stylized humanoid structure, a choice intended to enhance the perception of the pet as a ´´social partner" rather than a mere digital icon.

Species-specific traits were preserved to ensure character recognition: the lynx features pointed ears with dark tufts and elongated facial fur, while the panda emphasizes rounded, high-contrast features. Surface coloring was achieved through UV unwrapping and texture painting, prioritizing a clean, stylized aesthetic over high-fidelity realism to ensure visual consistency across different mobile display environments and to minimize system overhead during real-time notification delivery.

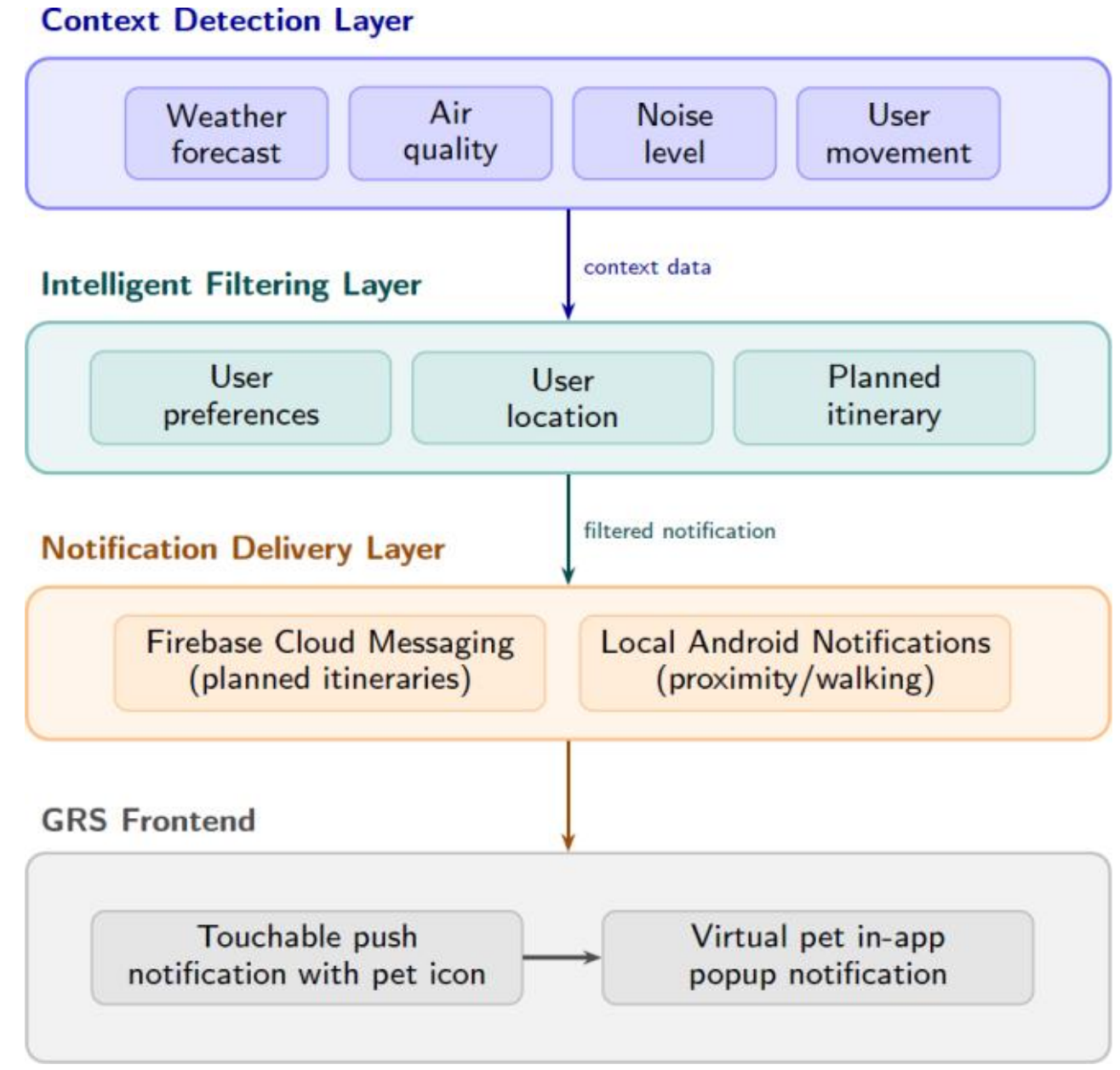

*Figure 2: Three-layer architecture of the context-aware notification system.*

For this proof-of-concept, the virtual pets were implemented with a consistent positive and friendly facial expression. While the architecture supports future dynamic animations, which are used in the embedded Virtual Pet Game (see [Alves et al. 2025]), the static "happy" state for the notifications was intentionally maintained during the pilot study to evaluate if a consistently welcoming companion could mitigate the negative emotional impact of adverse environmental alerts (e.g., air quality or weather warnings). This design choice follows the principle that a stable, non-threatening social proxy can reduce user stress and increase receptiveness toward corrective system advice.

### 3.3 Virtual pet frontend integration

The virtual pet assets were exported from Blender as high-resolution PNG images to ensure visual fidelity and transparency across diverse mobile display environments. This approach allows for a consistent representation of the mascots while minimizing the computational overhead during real-time notification delivery.

The virtual pet was integrated into the notification interface as a central communication agent. When a contextual alert is triggered, a push notification along with the pet image is presented in the mobile device status bar (Figure 3b), that when tapped, redirects to the application, which then retrieves the user's specific pet from the User Management Microservice (more details on the services in [Alves et al. 2024]). The pet is then presented in a centralized interactive dialog box, accompanied by a speech bubble (Figure 3(c-d)). This design simulates a direct conversation between the pet and the user, framing the recommendation as companion advice rather than a generic system alert.

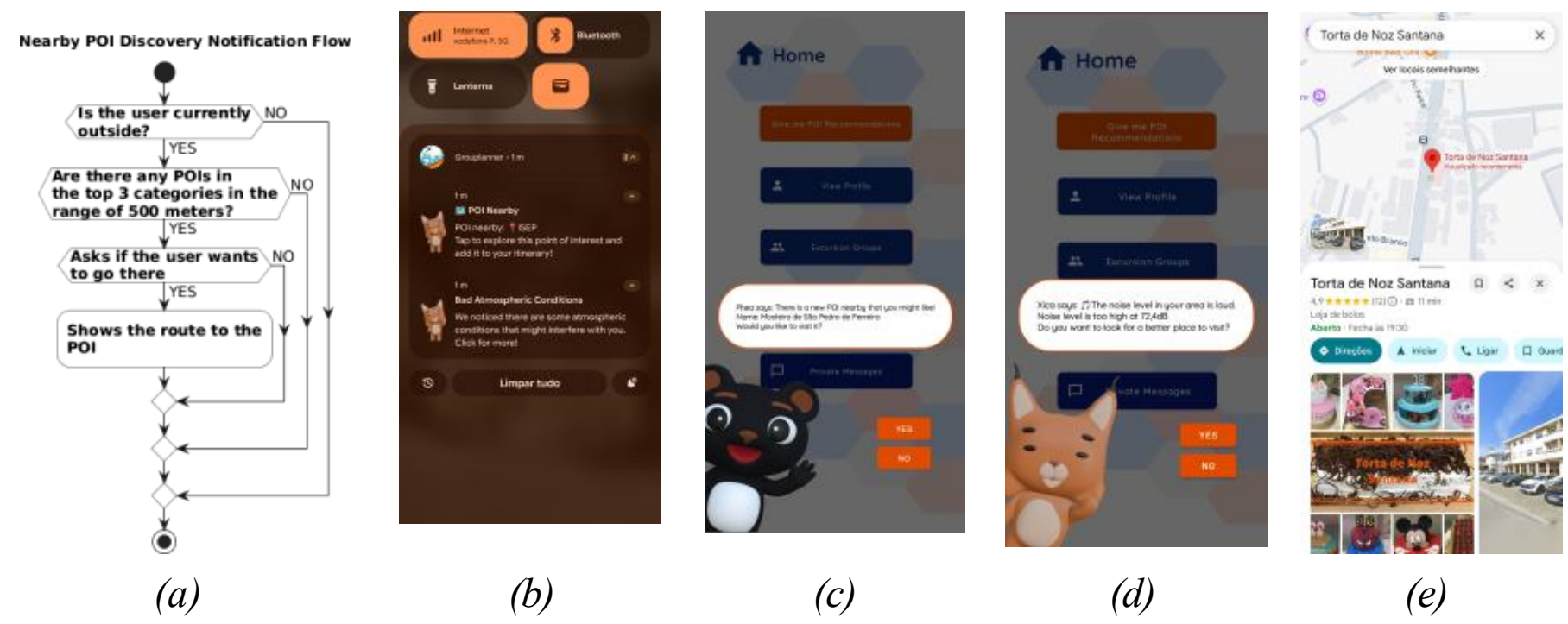

*Figure 3: (a) Nearby POI discovery flow for proximity-based recommendation delivery; (b) Push notifications for a nearby POI and bad weather conditions; (c) Virtual pet notification for a nearby POI; (d) Virtual pet notification for prejudicial noise level; (e) Google maps interface with a suggested POI for shelter.*

From a technical perspective, the interface utilizes the Android's native dialog system to ensure that notifications appear prominently over the current application state without requiring complex system permissions. This ensures a seamless user experience across different Android versions. The visual layout is standardized to maintain a consistent presence, regardless of whether the alert concerns a weather warning or a proximity-based POI suggestion. This stable and familiar presence is intended to foster a sense of social proxying, making the app's feedback feel more interactive and enjoyable for the user.

### 3.4 Notification architecture and contextual scenarios

The system implements three distinct context-aware notification scenarios, each utilizing specific trigger mechanisms and data sources. All notifications are driven by the underlying personality-based algorithm [Alves et al. 2024], which leverages the Big Five personality dimensions to tailor content to the user's psychological profile (Openness, Conscientiousness, Extroversion, Agreeableness, and Neuroticism), travel-related preferences & concerns, and specific constraints (e.g., fears/phobias or disabilities). For instance, users scoring high in Openness receive suggestions for cultural venues, while those high in Extraversion are prompted with social events (more details in [Alves et al. 2023]).

From a user experience (UX) perspective, the delivery follows a layered approach designed to balance visibility with minimal intrusiveness. As previously mentioned, alerts initially appear as standard Android status bar notifications; upon interaction, they expand into interactive in-app popups featuring the virtual pet companion. These popups utilize character-mediated speech bubbles to provide clear justifications for each alert, helping the user understand the rationale behind a specific recommendation or warning.

As illustrated in Figure 3a, Figure 5, and Figure 4, notifications are triggered in three scenarios:

– **Scenario 1 - Proximity-based Recommendations:** Triggered when matching POIs are identified near the user's real-time location;

- **Scenario 2 - Real-time Environmental Alerts:** Triggered by adverse shifts in atmospheric conditions - such as prejudicial air quality, high noise levels, or sudden precipitation - affecting the user's current surroundings;
- **Scenario 3 - Excursion Forecast Alerts:** Triggered for planned individual or group activities scheduled within a five-day window, warning of potential adverse weather at the destination.

To ensure safety and relevance, proximity and real-time environmental alerts require active background location tracking and only function when the user is moving at walking speeds. If the system detects high-speed movement (e.g., traveling in a vehicle), notifications are automatically suppressed to avoid unnecessary distraction.

#### 3.4.1 Environmental thresholds and severity assessment

To evaluate environmental safety, the system distills complex sensor data into actionable alerts. For air quality, we utilized the EPA's Air Quality Index (AQI) [AirNow 2025, EPA 2024], which monitors five main air pollutants: ground-level ozone ($O_3$), particle pollution (PM2.5 and PM10), carbon monoxide ($CO$), sulfur dioxide ($SO_2$), and nitrogen dioxide ($NO_2$) (see Table 1). The AQI uses six color-coded categories corresponding to a certain range level. Higher AQI values indicate worse air pollution and greater health risk. For instance, values of 50 or below (green category) are classified as good air quality with low or no implications to health, while values above 300 (maroon category) are classified as hazardous (see Table 2). To ensure the virtual pet's communication remains concise and less cognitively demanding, we simplified the six standard AQI categories into a binary classification: Healthy (AQI $\leq 50$) and Unhealthy (AQI $> 50$, covering sensitive and general populations). This ensures that the character-mediated notifications provide clear companion advice without overwhelming the user with technical pollutant concentrations.

| Pollutant | Unhealthy Values |
|---|---|
| PM2.5 | $> 35\ \mu g/m^3$ |
| PM10 | $> 150\ \mu g/m^3$ |
| $NO_2$ | $> 100$ ppb |
| $O_3$ | $> 120$ ppb |
| CO | $> 9$ ppm |

*Table 1: Unhealthy concentration thresholds for each air pollutant, adapted from EPA NAAQS standards [EPA 2024]*

Regarding urban noise, the system utilizes the Equivalent Continuous Sound Level ($L_{eq}$), also referred to as the equivalent continuous sound pressure level, which represents the average sound exposure over time, making it widely used in environmental, urban, and occupational acoustics [SVANTEK 2025, Coelho and Ferreira 2009]. When measured with A-weighting (the range of sound perceivable by the human ear [Lisboa E-Nova 2020]), it is typically expressed as $L_{Aeq}$ or $dB(A)$ $L_{eq}$, which approximates the frequency sensitivity of human hearing. Following cited daytime outdoor comfort thresholds and regulatory limits for residential areas or mixed-use areas [WHO 2018, República 2007, Lisboa E-Nova 2020], we adopted 55 $dB(A)$ $L_{eq}$ as the reference threshold for typical urban walking scenarios. Noise levels exceeding this value are assessed as harmful or prejudicial to the exploration experience, triggering a pet-mediated warning that suggests quieter alternative routes or indoor activities (see Figure 3b,d).

| AQI Values | AQI Color | Category | Health Implications |
|---|---|---|---|
| 0–50 | Green | **Good** | Satisfactory air quality. No or low health implications. |
| 51–100 | Yellow | **Moderate** | Acceptable air quality. Some pollutants may slightly affect very few hypersensitive individuals. |
| 101–150 | Orange | **Unhealthy for Sensitive Groups** | Sensitive individuals may be affected. |
| 151–200 | Red | **Unhealthy** | Sensitive individuals may experience more serious conditions. The hearts and respiratory systems of healthy people may be affected. |
| 201–300 | Purple | **Very Unhealthy** | Considered as an health alert. Everyone, including healthy people, are at risk of having health effects. |
| >= 301 | Maroon | **Hazardous** | Emergency health warning. Everyone, including healthy people, are more likely to have health effects. |

*Table 2: The six Air Quality Index categories provided by EPA, adapted from [AirNow 2025].*

Finally, precipitation is monitored to prevent disruptions to tourist itineraries. Based on IPMA's and World Meteorological Organization's (WMO) standards [IPMA 2021b, WMO 2024], the system tracks four intensity categories: moderate rain, heavy rain, intermittent rain, intermittent heavy rain (see Table 3). Assuming that light rain would not be considered an impediment to travel/visit a location, notifications are only emitted for conditions that significantly impact travel comfort, ensuring that the virtual pet only interrupts the user for meaningful contextual shifts.

| Precipitation (mm/h) | Category | Travel Safety |
|---|---|---|
| 0.0 | No rain | Safe |
| [0.0, 2.5[ | Light rain | Safe |
| [2.5, 10.0[ | Moderate rain | Unsafe |
| [10.0, 50.0[ | Heavy rain | Unsafe |
| >50.0 | Violent rain | Unsafe |

*Table 3: Rainfall intensity categories and travel safety, adapted from WMO/MANOBS standards [WMO 2024]*

### 3.4.2 Scenario 1: proximity-based recommendations

To provide contextually relevant suggestions during active exploration, the system delivers proximity-based recommendations when the user is detected walking outdoors. This process, managed by the *Context Detection Layer*, utilizes the Android Activity Recognition API to monitor movement patterns. To ensure the user is in an "exploration state" and to mitigate unwanted interruptions, a temporal threshold was implemented: the system only triggers a recommendation after detecting five consecutive minutes of walking (configurable).

Once this threshold is met, the system retrieves the user's profile from the MAMS, via HTTPS REST, and requests a recommendation from the REMS based on the current GPS coordinates. The engine identifies matching POIs within a configurable 500-meter radius. The resulting notification is then delivered through the virtual pet interface, framing the suggestion as a serendipitous discovery (see Figure 3b-c). To further respect the user's attentional state, the detection timer resets if the user remains stationary for more than one minute (e.g., during a rest period), preventing notifications when the user is not actively exploring. The proximity-based recommendation flow is illustrated in Figure 3a.

### 3.4.3 Scenario 2: Real-time environmental alerts

When the user is detected walking outdoors, the *Context Detection Layer* periodically polls (every 10 minutes, configurable) real-time IoT sensor data from the Porto Digital FIWARE platform. This smart-city infrastructure monitors air quality, noise levels, and precipitation via a distributed network of over 150 sensors across Porto, Portugal. Porto Digital [Porto Digital n.d.a] is a non-profit association created in 2004 focused on digital innovation and smart city initiatives in Porto, Portugal. It plays a key role in urban platforms, IoT infrastructure, and public digital services, having developed a smart city concept in the city of Porto, where over 100 air quality sensors and 50 noise level sensors are placed throughout the city.

The system interfaces with Porto Digital FIWARE Broker API [Porto Digital n.d.b] using the Next Generation Service Interfaces - Linked Data (NGSI-LD) protocol to retrieve three types of environmental observations:

- **Air Quality:** Concentrations of particles superior to 2.5 $\mu g/m^3$ and 10 $\mu g/m^3$ (PM2.5 and PM10, respectively), and $NO_2$ (nitrogen dioxide), $CO$ (carbon monoxide), and $O_3$ (ozone) atmospheric gases;
- **Noise Levels:** Acoustic data measured in decibels ($dB$);
- **Precipitation:** Real-time readings in $mm/h$ from local weather stations.

To provide personalized decision support, the system utilizes the Haversine formula to identify the sensor closest to the user's GPS coordinates. These readings are then evaluated against severity thresholds based on Environmental Protection Agency's (EPA) guidelines and comfort levels [AirNow 2025, Noise Awareness Day 2026] (more details in Section 3.4.1). If unsafe conditions are detected (e.g., PM2.5 $> 25\mu g/m^3$ or noise $> 55dB(A)$), the system triggers a layered notification flow (see Figure 4). First, a push notification alerts the user of the adverse conditions (see Figure 3b). Upon interaction, the virtual pet popup appears, asking if the user wishes to seek a safer location (see Figure 3d). If affirmative, the system communicates with the MAMS to determine the user's preferences and then asks REMS for indoor POI suggestions - within a 500m radius for rain, or further for air/noise pollutants - prioritizing the user's preferences if possible, otherwise the closest random indoor POI, and provides direct navigation via Google Maps (see Figure 3e). If the system detects that there are no possible POI available, it asks the user to walk a bit more until the system can find a proper place. By mediating this technical alert through the virtual pet, the framework aims to reduce the alarmist nature of environmental warnings, transforming them into "companion-led" safety advice that encourages proactive decision-making.

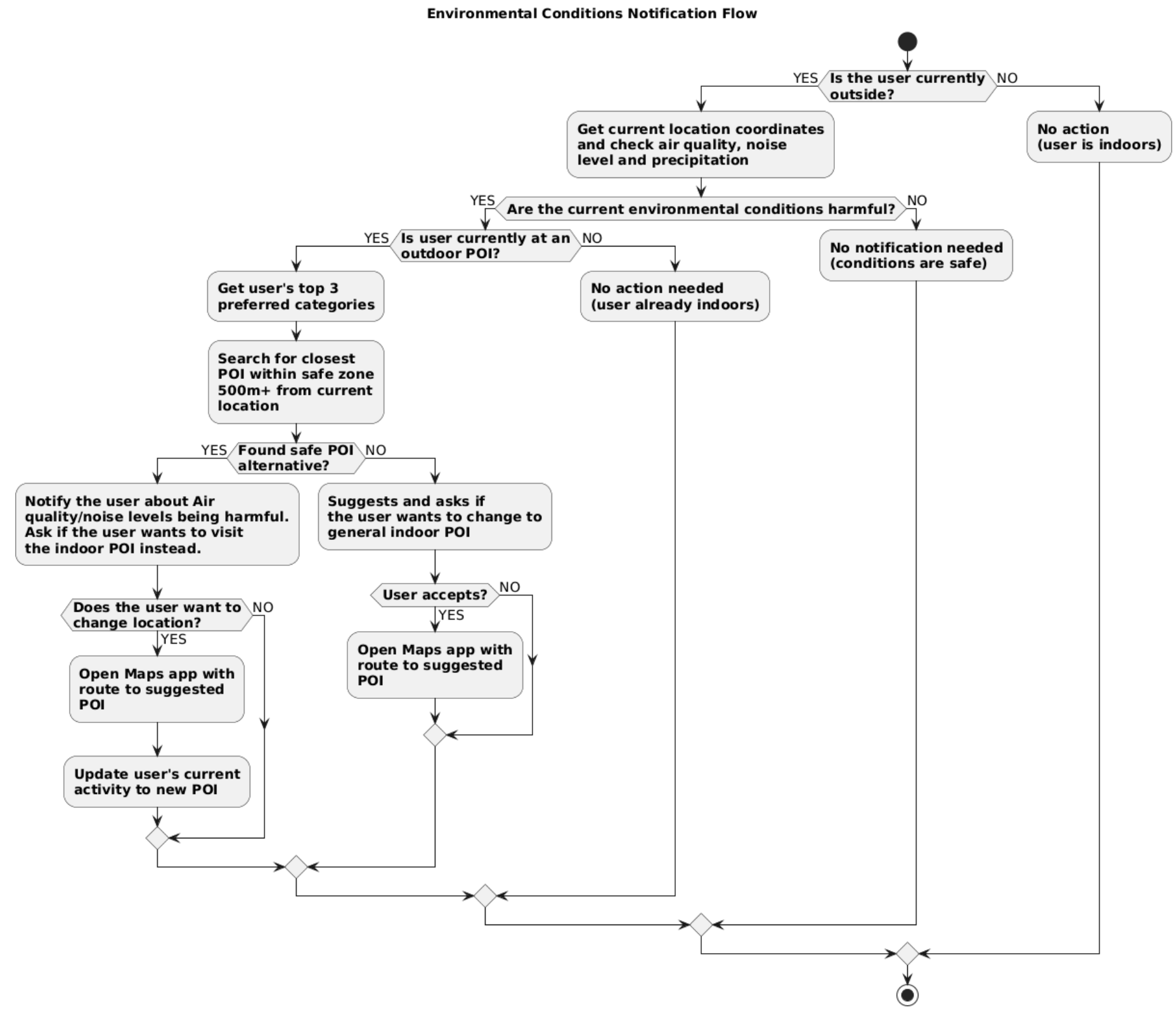

*Figure 4: Notifications flow for real-time environmental alerts (prejudicial air, noise, and/or precipitation) during outdoor walking.*

### 3.4.4 Scenario 3: excursion forecast alerts

The system also monitors weather forecasts for scheduled individual or group excursions within a 5-day window. The *Context Detection Layer* polls the Instituto Português do Mar e da Atmosfera (IPMA) [IPMA n.d.] API every 24 hours (configurable) to retrieve official meteorological data, including precipitation probabilities, wind speeds, temperature ranges, and weather type classifications [IPMA 2021b]. The weather type classifications are based on IPMA standards and encompass weather types like heavy rain, storms, snow, fog, and extreme temperatures, among others. The MAMS dynamically maps excursion POI locations to IPMA district codes to ensure spatial accuracy in its assessments [IPMA 2021a].

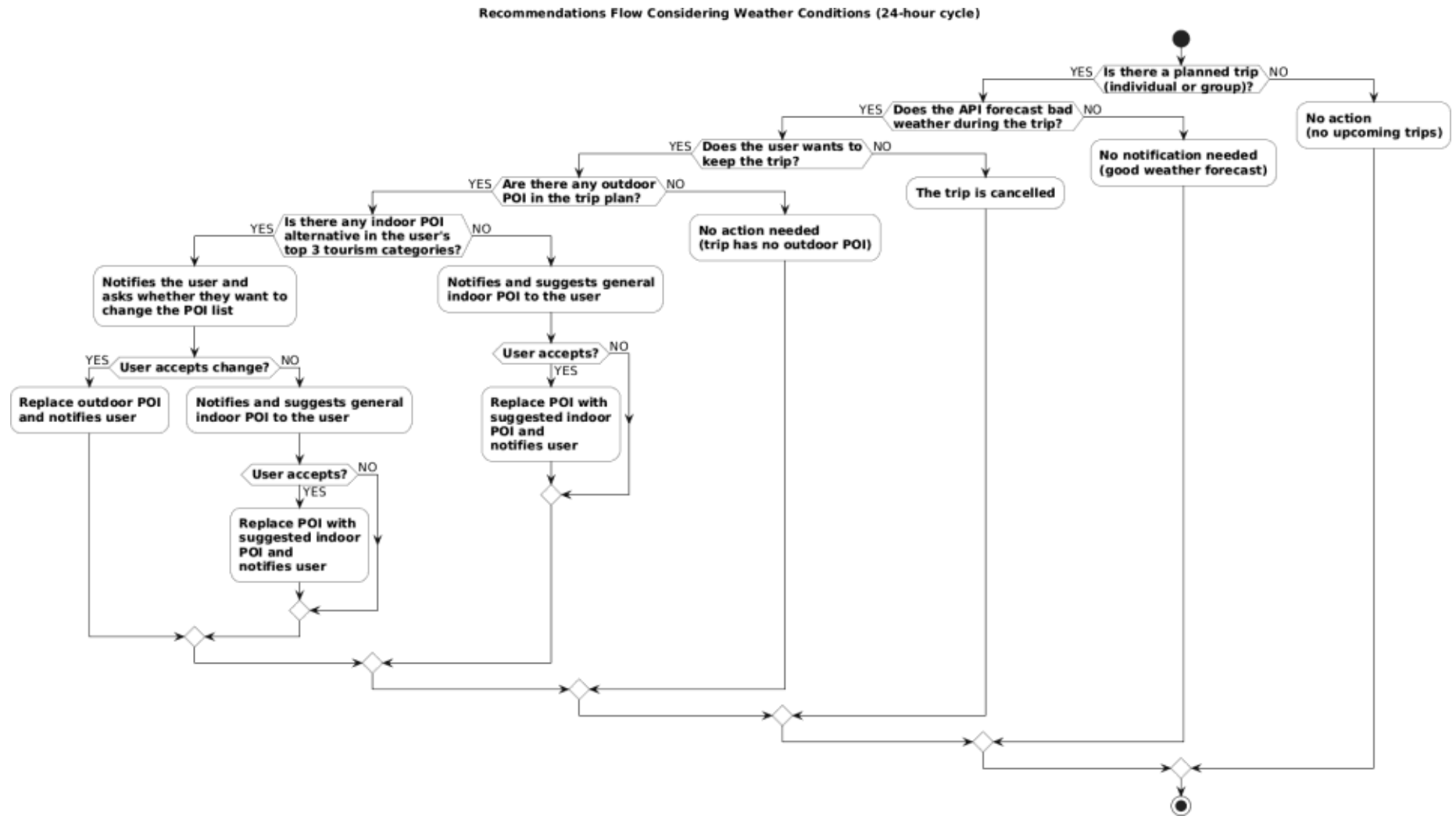

*Figure 5: Weather alert notifications flow showing the end-to-end process from forecast detection to virtual pet popup display.*

The MAMS evaluates these forecasts against standard safety thresholds to determine if a scheduled trip is compromised (see Appendix Table A.1). If adverse weather is detected (e.g., heavy rain, extreme temperatures, strong wind), the system initiates a proactive recommendation loop: the MAMS fetches the user's or group members' tourism categories preference along with a list of POI that should be included or excluded (see [Alves et al. 2024]), and sends the information to the REMS, which then identifies alternative indoor POIs that match the tourist(s)' preferences (or general indoor POI if none available, see Figure 5), replacing weather-sensitive outdoor activities.

Alert delivery is managed through Firebase Cloud Messaging, utilizing unique Firebase tokens stored in the User Management microservice database to ensure high-reliability push notifications. Each notification is assigned a weather severity level (LOW, MEDIUM, HIGH, or CRITICAL) derived from the meteorological readings (detailed in Appendix Table A.1) and sent to the user. By providing these alerts up to five days in advance, the framework acts as a predictive assistant, allowing users to adjust their plans through the virtual pet popup interface, which provides the necessary context and justifications for the suggested changes.

### 3.5 Pilot study design and procedure

To evaluate the feasibility of the proposed approach, a counterbalanced within-subjects experimental design was employed, comparing two versions of the framework: (i) a baseline version with contextual notifications without the virtual pet-mediation and (ii) a full-featured version incorporating the pet-mediated contextual notification mechanism [1]. The independent variable was the notification delivery method (standard system vs. pet-mediated), and each participant was exposed to both conditions for a period of two weeks per version.

Participants ($n$ = 11) were recruited through oral dissemination and informal invitations within a research center environment. While the sample is relatively small and predominantly composed of individuals with a technology-oriented background, it is appropriate for a preliminary proof-of-concept evaluation. This homogeneity allowed for a focused assessment of the system's technical feasibility and the baseline acceptance of the pet-mediated interaction design. Furthermore, the within-subjects design ensured that each participant served as their own control, reducing error variance associated with individual differences in RS familiarity or digital literacy.

To mitigate potential order effects - such as learning, habituation, and carryover influences - two exposure sequences were implemented. Participants in Sequence A utilized the baseline version first, followed by the full-featured version, while those in Sequence B followed the reverse order. This counterbalancing was critical to preserving internal validity during the pilot evaluation. Prior to system interaction, participants completed an online pre-questionnaire for characterization and received standardized instructions for installing the GRS application, registration, and guidance on core functionalities to ensure procedural consistency. During system onboarding, they completed the embedded Big Five Inventory (BFI) in the profile menu to enable personality-based recommendations, addressing the cold-start problem [Alves et al. 2024].

Participants were instructed to explore each version freely within the northern region of Portugal. The protocol required requesting at least one individual recommendation and walking within Porto to allow the GRS to trigger context-aware notifications under realistic mobility conditions. While the framework supports group recommendations, this study prioritized individual perception to evaluate the virtual pet as a communication mediator before its deployment in collaborative scenarios. After completing both two-week phases, participants filled in a comprehensive post-questionnaire, including the UEQ-S scale [Hinderks 2017], which is the short version of the User Experience Questionnaire (UEQ), comprised of 8-items on a 7-point Likert scale, designed to measure hedonic and pragmatic quality; specific items regarding notification utility, acceptance and pet-mediation were analyzed by semi-structured interviews and open-ended qualitative questions.

## 4 Results and Discussion

### 4.1 Participants characterization

The pilot study involved 11 participants, whose demographic and background details are summarized in Table 4. The group was aged between 22 and 43 years ($M$ = 30.0, $SD$ = 7.0) and was composed of 8 males (72.7%) and 3 females (27.3%). The sample exhibited high educational levels, with all participants holding a university degree (63.6% at the Master's level) and a strong technical orientation, as 90.9% had a background in Engineering and Technology.

Regarding their familiarity with the study's core domains, all participants reported previous experience using commercial tourism RS. Furthermore, the majority (63.6%) were active mobile game players, while the remaining participants (36.4%) had prior experience with mobile gaming. This profile represents a group of "expert users" with high digital literacy, which is particularly suitable for a preliminary proof-of-concept focused on assessing the technical feasibility and interaction logic of context-aware notifications.

[1] The full version APK (Android only) can be downloaded from https://www.gecad.isep.ipp.pt/grouplanner/dissemination.html.

| Category | Option | $n$ | % |
|---|---|---|---|
| Age | Mean ± SD (range) | 30.0 ± 7.0 (22–43) | |
| Gender | Male | 8 | 72.7 |
| | Female | 3 | 27.3 |
| Education | Master's degree | 7 | 63.6 |
| | Bachelor's degree | 3 | 27.3 |
| | Doctorate | 1 | 9.1 |
| Field of study | Engineering & Technology | 10 | 90.9 |
| | Medical & Health Sciences | 1 | 9.1 |
| Mobile gaming | Not currently | 4 | 36.4 |
| | Once a week | 3 | 27.3 |
| | Several times a day | 2 | 18.2 |
| | Several times a week | 2 | 18.2 |
| Tourism RS used[a] | Booking.com | 7 | 63.6 |
| | Tripadvisor | 5 | 45.5 |
| | Airbnb | 2 | 18.2 |
| | Facebook Events | 2 | 18.2 |
| | Google Maps GP | 1 | 9.1 |
| | TheFork | 1 | 9.1 |

[a]The total percentage surpasses 100% as multiple answers were allowed.

*Table 4: Participants characterization ($n$ = 11).*

### 4.2 Pilot study results

To evaluate the feasibility and compare the perceived user experience between the baseline and pet-mediated versions, a paired non-parametric test was conducted using the Wilcoxon signed-rank test (two-tailed, $\alpha = 0.05$). Effect sizes were estimated using the matched-pairs rank-biserial correlation ($r_{rb}$), which is specifically suited for small samples ($n = 11$) and ordinal Likert-type data. $|r_{rb}|$ was interpreted as: negligible <0.10, small 0.10–0.29, medium 0.30–0.49, large ≥0.50. Given the pilot nature of this proof-of-concept, results are interpreted as exploratory and directional, focusing on identifying meaningful trends in user engagement and motivational response rather than providing generalizable confirmatory evidence. The detailed statistical results are presented in Appendix A.

#### 4.2.1 Impact on user experience

The analysis of the UEQ-S scale revealed a positive descriptive shift in user experience when notifications were mediated by the virtual pet (see Figure 6 and Appendix Table A.2). The overall UEQ-S mean improved from $M = 4.68$ ($SD = 0.95$) in the baseline to $M = 4.93$ ($SD = 0.91$) in the pet-mediated version. Although this difference did not reach conventional statistical significance ($W = 9.5$, $p = .074$), the medium-to-large effect size ($r_{rb} = .655$) indicates a promising trend regarding the companion's ability to enhance the interaction quality.

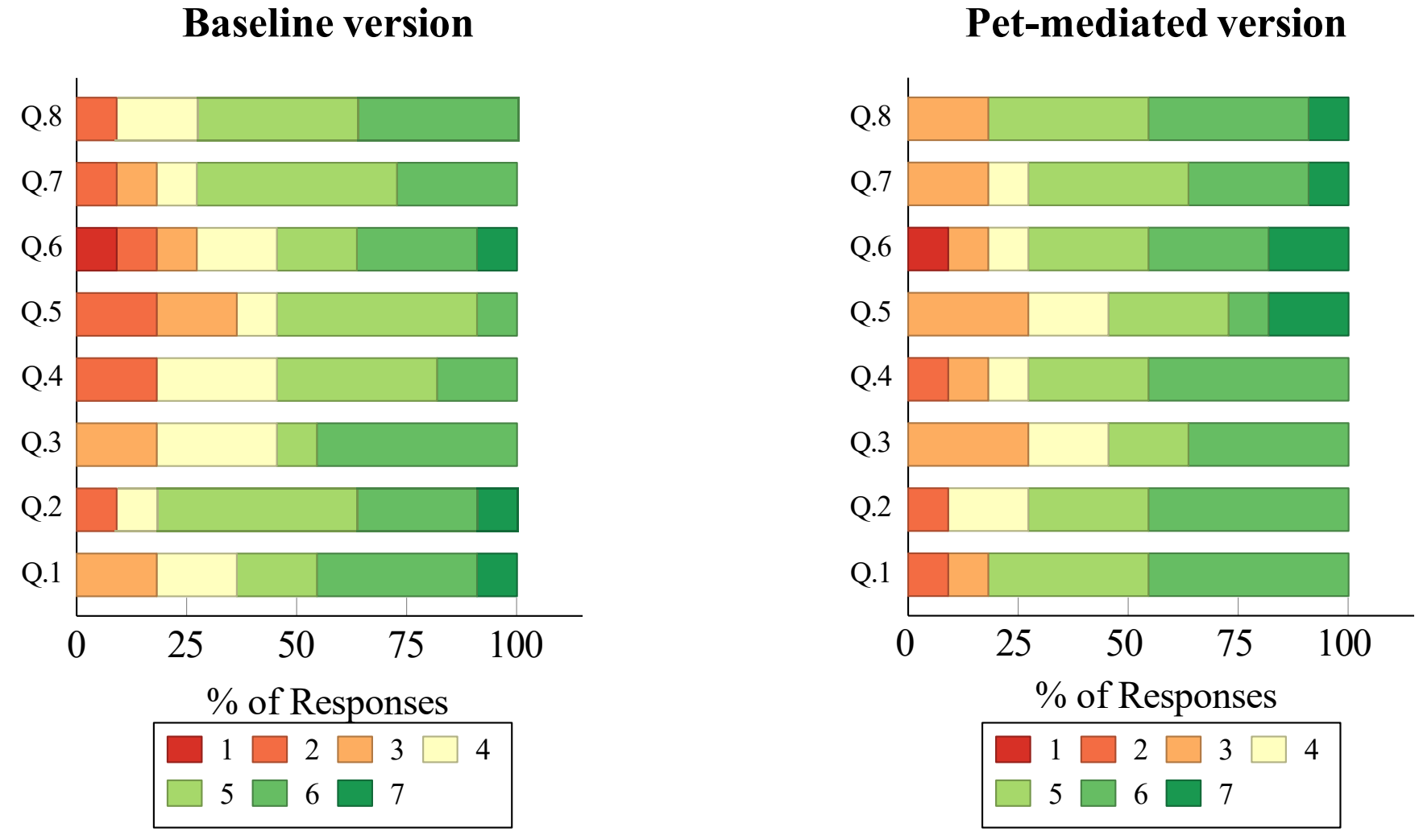

*Figure 6: Aggregated responses (%, $n = 11$) to the UEQ-S questionnaire for the baseline (left) and pet-mediated (right) versions. Scale: 1 (negative pole) to 7 (positive pole). Q.1 Obstructive–Helpful, Q.2 Complicated–Easy, Q.3 Inefficient–Efficient, Q.4 Confusing–Clear, Q.5 Boring–Exciting, Q.6 Uninteresting–Interesting, Q.7 Conventional–Inventive, Q.8 Usual–Innovative. Items Q.1–Q.4 measure Pragmatic Quality (PQ); items Q.5–Q.8 measure Hedonic Quality (HQ).*

This improvement was primarily driven by the *Hedonic Quality* dimension, which rose from $M = 4.55$ ($SD = 1.20$) to $M = 4.98$ ($SD = 1.17$), yielding a medium-to-large effect ($W = 9.0$, $p = .070$, $r_{rb} = .673$). Conversely, the *Pragmatic Quality* remained stable across both conditions (baseline: $M = 4.82$; pet-mediated: $M = 4.89$, $W = 16.5$, $p = .613$, $r_{rb} = .267$). This stability is a significant result for the proof-of-concept, as it demonstrates that adding a virtual pet layer to the communication process does not hinder the system's efficiency or introduce unnecessary complexity for the user.

Item-level analysis further clarified these trends. The most notable improvement was found in the "Confusing–Clear" pair ($W = 0.0$, $p = .063$, $r_{rb} = 1.00$), suggesting that the pet-mediated speech bubbles helped clarify the notification rationale. Positive shifts were also observed in hedonic attributes such as "Boring–Exciting" ($r_{rb} = .607$) and "Uninteresting–Interesting" ($r_{rb} = .611$). Given the exploratory nature of this pilot study, the combination of high effect sizes and non-significant $p$-values suggests that the study is likely underpowered rather than lacking an effect. These findings suggest that virtual companions can humanize the notification process, increasing stimulation and interest while maintaining the pragmatic utility of the RS interface.

#### 4.2.2 User perceptions of pet-mediated notifications

A quantitative and qualitative analysis on the pet-mediated notification interface was conducted, using semi-structured interviews and manually recording the responses to the items mentioned in Section 3.5 (see Figure 7 and Appendix Table A.3). The overall average score of $M = 5.17$ ($SD = 0.67$) indicates a high level of acceptance for the proof-of-concept. All evaluated items scored above the midpoint of 4.5, suggesting that the integration of the virtual companion was perceived as a positive addition to the recommendation process.

The *Utility* (U) sub-scale achieved the highest average ($M = 5.25, SD = 0.75$), with participants reporting a clear understanding of the notification motives (Q13.7, $M = 5.36, SD = 0.64$) and appreciating the clarity of the justifications provided through the pet's speech bubbles (Q13.3, $M = 5.36, SD = 0.77$). These results suggest that character-mediated communication can support transparency, helping users understand the rationale behind contextual alerts.

Regarding *Acceptance* (A), notifications were considered generally welcome (Q13.2, $M = 5.36, SD = 0.88$) and timely (Q13.10, $M = 5.27, SD = 0.86$). While '′Intrusiveness" (Q13.6) remains a common challenge in mobile RS, the score of $M = 4.82, SD = 1.34$ suggests that the virtual pet managed to mitigate this effect, making the alerts feel less bothersome. This is further supported by the *VP mediation* (VP) sub-scale, where participants agreed that the pet made notifications more pleasant (Q13.4, $M = 5.36, SD = 1.15$). However, item Q13.11 ("Pet made communication more natural") showed higher variability ($M = 4.91, SD = 1.44$), reflecting the diverse user orientations identified in the qualitative feedback.

The analysis of open-ended responses (Q17–Q19) and of the interviews provides a deeper understanding of the ′′why" behind the quantitative trends, revealing a clear segmentation in user expectations and interaction styles.

Regarding the most liked aspects (Q17), participants (*Pi*) highlighted three pillars of the system:

- **High-Quality Personalization:** Four participants ($P1$, $P4$, $P6$, $P8$) explicitly valued the alignment between personality traits and recommendations. $P8$ noted that the "integration of personality questions" was a specific value-add.
- **Contextual Utility:** The proximity-based recommendations were praised, with $P7$ mentioning it helps discover "hidden gems" and $P10$ highlighting the utility of weather alerts.
- **Gamified Engagement:** For users with a hedonic orientation, the pet was a central strength. $P8$ described the interaction as ′′fun and immersive," while $P9$ found that the pet made the overall experience "much more interesting."

The feedback on less-liked aspects (Q18) identified some critical refinements, where despite the pet mediation, $P9$ noted that alerts could still be "persistent" or appear at "inopportune times," highlighting the challenge of perfect temporal context.

The influence of the virtual pet (Q19) emphasized the Hedonic vs. Pragmatic differentiation. The Hedonic group ($P1$, $P2$, $P4$, $P8$, $P9$) described the pet as a driver of "proximity," "involvement," and "immersion." In contrast, the Pragmatic group ($P3$, $P7$, $P11$) remained indifferent or critical, with $P7$ stating that for a recommendation purpose, "having a pet or not is indifferent." Interestingly, $P10$ and $P11$ reported a "negative influence," directly questioning the conceptual fit of a pet within a tourism tool.

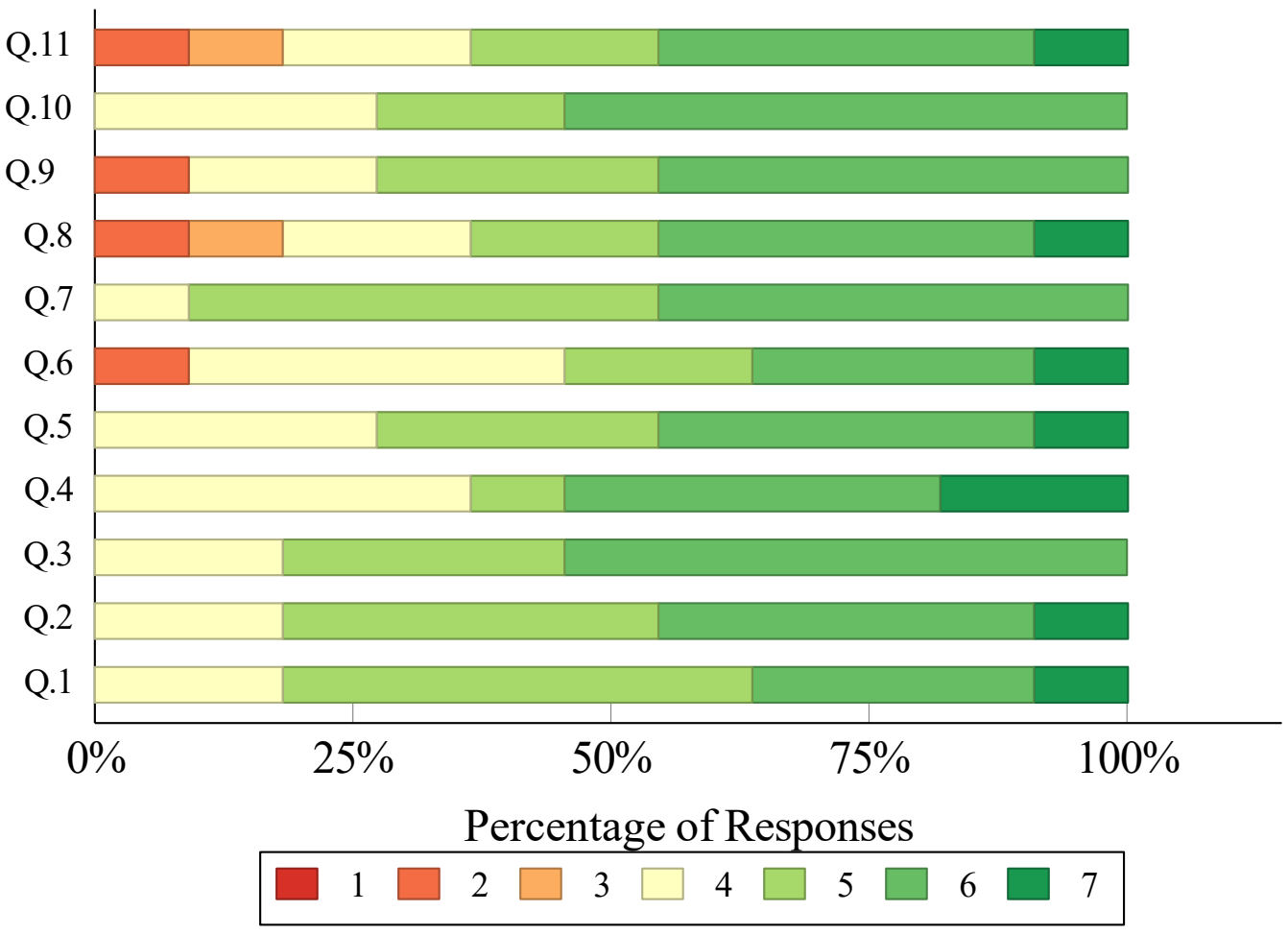

*Figure 7: Aggregated responses to the Pet-mediated notifications scale ($n = 11$, 1–7). Q.1, Q.3, Q.5, Q.7, Q.9: Utility; Q.2, Q.6, Q.10: Acceptance; Q.4, Q.8, Q.11: Virtual Pet mediation.*

These qualitative results triangulate the UEQ-S findings: gamification effectively boosts the *Hedonic Quality* for experience-seekers but can hinder the *Pragmatic Quality* for efficiency-oriented users. This "conceptual mismatch" reported by $P3$ and $P7$ provides a strong design justification for adaptive interfaces. Future iterations of this proof-of-concept should allow users to toggle the gamification layer, ensuring that the system provides "companion-led advice" for those who value empathy, and a "minimalist tool" for those focusing solely on utility.

## 5 Conclusions and future directions

This paper presented a proof-of-concept for a context-aware notification framework where real-time environmental data - including air quality, noise, and weather - and proximity-based POI recommendations are mediated by a virtual pet companion. The study investigated whether anthropomorphic mediation could mitigate notification fatigue and improve the acceptance of safety-critical alerts by transforming technical data into "companion-led" advice.

The exploratory within-subjects pilot study ($n = 11$) suggests the feasibility and high user acceptance of this approach, with several key findings: (1) The virtual pet effectively acted as a social proxy, making notifications feel significantly more pleasant and welcome. This suggests that character-mediated communication can soften the perceived intrusiveness of real-time alerts, fostering a more positive interaction environment; (2) The system served as an Explainable AI interface (XAI). By providing clear, character-driven justifications (high scores in Q13.3 and Q13.7), the pet helped users understand the rationale behind environmental shifts, directly supporting proactive decision-making; (3) UEQ-S results confirmed that the virtual pet layer significantly enhanced hedonic quality (stimulation and novelty) without compromising the system's pragmatic utility.

This suggests that gamified mediation can be integrated into functional tools without introducing unnecessary cognitive overhead.

However, the study identified a clear polarization between user orientations. While hedonic-oriented users embraced the companion's social presence, pragmatic users prioritized immediate efficiency and occasionally perceived the virtual pet as a source of minor cognitive friction. This highlights the necessity for adaptive notification strategies that allow users to calibrate the level of social mediation based on their personal preferences.

Based on these preliminary findings, future work will focus on implementing dynamic emotional animations (e.g., a "worried" pet during hazardous air quality alerts) to evaluate the impact of increased social cues; conducting a study with a larger sample (at least $n > 30$) to examine how long-term companion bonding affects sustained engagement and response rates; exploring deeper XAI-driven dialogues where the pet can offer more detailed explanations for its advice upon user request; and developing an "opt-in" gamification model that streamlines the interface for pragmatic users while maintaining the full experience for those seeking hedonic engagement.

This study provides initial evidence that virtual pets are effective, non-intrusive mediators for context-aware communication, successfully humanizing technical data and fostering a more supportive tourism experience.

**Acknowledgements**

This work was supported by the PRR-ATT Project, reference C645192610-00000060, under the Portuguese Recovery and Resilience Plan (PRR) and NextGenerationEU European Funds, and by national funds through Fundação para a Ciência e a Tecnologia (FCT) under the Project UID/00760/2025. The authors would like to thank all the participants in the pilot study and for their useful insights.

## Ethics statement

This work involved human subjects in its research. Approval of all ethical and experimental procedures and protocols were granted by the Polytechnic of Porto Data Protection Office, and performed in line with the institution's requirements. The questionnaires used were anonymous and confidential, and the collected data was analyzed in aggregated form. All participants signed an informed consent to participate in the study, ensuring confidentiality, anonymity, and informing them of the aims of the research and where the data was stored.

## Disclosure statement

The authors report there are no competing interests to declare.

## Declaration of generative AI and AI-assisted technologies in the manuscript preparation process

During the preparation of this work the authors used Claude in order to create the charts and tables structures, and to correctly format them according to the template guidelines.

Beside the usual scientific databases, Perplexity was used to find relevant literature to aid in the state of the art review. Gemini was used to improve text language and readability. After using these tools, the authors reviewed and edited the content as needed and take full responsibility for the content of the published article.

# A Auxiliary data and statistics

| Severity | Precipitation | Wind Speed | Temperature | Weather Type |
|---|---|---|---|---|
| **LOW** | < 2.5 mm/h | < 30 km/h | 5–30°C | Light rain, Cloudy, Light snow |
| **MEDIUM** | 2.5–10 mm/h | 30–50 km/h | 0–5°C or > 30°C | Moderate rain, Snow, Light fog |
| **HIGH** | 10–25 mm/h | 50–80 km/h | < 0°C or > 35°C | Heavy rain, Heavy snow, Dense fog |
| **CRITICAL** | > 25 mm/h | > 80 km/h | < −10°C or > 40°C | Storms, Extreme winds, Extreme temperatures |

*Table A.1: Weather severity level mapping to meteorological conditions.*

| Measure | Baseline M ± SD | Pet-mediated M ± SD | $W$ | $p$ | $r_{rb}$ |
|---|---|---|---|---|---|
| Q.1 Obstructive–Supportive | 5.00 ± 1.34 | 5.00 ± 1.10 | 5.0 | 1.000 | 0.000 |
| Q.2 Complicated–Easy | 5.09 ± 1.30 | 5.00 ± 1.34 | 6.0 | 1.000 | -0.200 |
| Q.3 Inefficient–Efficient | 4.82 ± 1.25 | 4.64 ± 1.21 | 4.0 | 1.000 | -0.200 |
| Q.4 Confusing–Clear | 4.36 ± 1.36 | 4.91 ± 1.04 | 0.0 | .063 | 1.000 |
| **Pragmatic Quality** | **4.82 ± 1.14** | **4.89 ± 0.99** | 16.5 | .613 | 0.267 |
| Q.5 Boring–Exciting | 4.09 ± 1.38 | 4.73 ± 1.49 | 5.5 | .172 | 0.607 |
| Q.6 Not interesting–Interesting | 4.45 ± 1.86 | 5.00 ± 1.34 | 7.0 | .172 | 0.611 |
| Q.7 Conventional–Inventive | 4.73 ± 1.27 | 5.00 ± 1.18 | 2.0 | .500 | 0.600 |
| Q.8 Usual–Leading | 4.91 ± 1.22 | 5.18 ± 1.25 | 3.0 | .375 | 0.600 |
| **Hedonic Quality** | **4.55 ± 1.20** | **4.98 ± 1.17** | 9.0 | .070 | 0.673 |
| **Overall UEQ-S** | **4.68 ± 0.95** | **4.93 ± 0.91** | 9.5 | .074 | 0.655 |

Items Q.1–Q.4 form the Pragmatic Quality subscale; Q.5–Q.8 form the Hedonic Quality subscale. $W$: Wilcoxon statistic; $p$: two-tailed $p$-value; $r_{rb}$: matched-pairs rank-biserial correlation (effect size); $|r_{rb}|$ interpreted as: negligible <0.10, small 0.10–0.29, medium 0.30–0.49, large ≥0.50.

*Table A.2: Descriptive statistics and Wilcoxon signed-rank test results for the UEQ-S (baseline vs. pet-mediated versions, $n = 11$, scale 1–7).*

| Item | Statement | M | SD |
|---|---|---|---|
| **Utility (U)** | | | |
| Q13.1 | Context notifications were useful | 5.27 | 0.86 |
| Q13.3 | The justifications were clear | 5.36 | 0.77 |
| Q13.5 | Increased recommendation relevance | 5.27 | 0.96 |
| Q13.7 | I understood the motive for each alert | 5.36 | 0.64 |
| Q13.9 | Helped me make decisions | 5.00 | 1.21 |
| *Sub-scale average* | | **5.25** | **0.75** |
| **Acceptance (A)** | | | |
| Q13.2 | The notifications were welcome | 5.36 | 0.88 |
| Q13.6 | The notifications were not intrusive | 4.82 | 1.34 |
| Q13.10 | The timing of the alerts was right | 5.27 | 0.86 |
| *Sub-scale average* | | **5.15** | **0.90** |
| **VP mediation (VP)** | | | |
| Q13.4 | The pet made notifications more pleasant | 5.36 | 1.15 |
| Q13.8 | Preferred pet-mediated over system alerts | 4.91 | 1.44 |
| Q13.11 | Communication felt more natural with the pet | 4.91 | 1.44 |
| *Sub-scale average* | | **5.06** | **1.10** |
| **Overall Mean Score** | | **5.17** | **0.67** |

*Table A.3: User perception of pet-mediated notifications, mean values (Q13,* $n = 11$*, scale 1–7). Sub-scales: Utility (U); Acceptance (A); Virtual Pet mediation (VP).*